\documentclass[11pt,twoside]{article}
\usepackage{./asp2014}
\usepackage{graphics, amsmath, amssymb, float}

\aspSuppressVolSlug
\resetcounters

\begin{document}

\title{Magnetic Field Diagnostics with Strong Chromospheric Lines}
\author{R.~Manso~Sainz,$^1$ T.~del~Pino~Alem\'an,$^{2}$ and R.~Casini$^2$
\affil{$^1$Max-Planck-Institut f\"ur Sonnensystemforschung, G\"ottingen, Germany \\
$^2$High Altitude Observatory, Boulder, CO, USA}
}

\paperauthor{Rafael~Manso~Sainz}{manso@mps.mpg.de}{}{Max-Planck-Institut f\"ur Sonnensystemforschung}{}{G\"ottingen}{Niedersachsen}{37077}{Germany}
\paperauthor{Tanaus\'u~del~Pino~Alem\'an}{tanausu@iac.es}{}{Instituto de Astrof\'\i sica de Canarias}{}{La Laguna}{Tenerife}{38205}{Spain}
\paperauthor{Roberto Casini}{casini@ucar.edu}{}{High Altitude Observatory}{}{Boulder}{CO}{80301}{USA}

\begin{abstract}
The complex spectropolarimetric patterns around strong chromospheric lines,
the result of subtle spectroscopic and transport mechanisms,
are sensitive, sometimes in unexpected ways, to the presence of magnetic fields in the 
chromosphere, which may be exploited for diagnostics.
We apply numerical polarization radiative transfer implementing 
partially coherent scattering by polarized multi-term atoms, in the presence of 
arbitrary magnetic fields, in planeparallel stellar atmospheres
to study a few important spectroscopic features: Mg~{\sc ii}~h-k doublet;
Ca~{\sc ii}~H-K doublet and IR triplet.
We confirm the importance of partial redistribution effects
in the formation of the Mg~{\sc ii}~h-k doublet in magnetized atmospheres, 
as previously pointed out for the non-magnetic case. 
Morevover, we show, numerically and analytically, that
a magnetic field produces measurable 
modications of the broadband linear polarization even for relatively small field strengths, 
while circular polarization remains well represented by the magnetograph formula.
We note that this phenomenon has already (unknowingly) been observed by UVSP/SMM, 
and the interest and possibility of its observation in stars other than the Sun.
The interplay between partial redistribution in the H-K doublet of Ca~{\sc ii} and 
metastable level polarization in its IR triplet allow diagnosing the chromospheric 
magnetic field at different layers and strengths. 
Our results suggest several new avenues to investigate empirically the magnetism
of the solar and stellar chromospheres.
\end{abstract}

Spectral lines that form in the chromosphere---a rarefied, 
relatively cool 
medium---correspond to strong resonant transitions
of abundant elements, 
and are 
scattered largely unaffected by collisions, 
which makes that even subtle, fragile radiation-matter interaction processes---partial
frequency redistribution (PRD) and coherence between incident and outgoing radiation 
\citep{Shine+75}, 
coherence between different atomic levels \citep{Stenflo80, Smitha+11, BelluzziTrujillo11}, 
atomic polarization in long-lived metastable levels \citep{MansoTrujillo03}---, become observable.
Magnetic fields, by removing the degeneracy of the atomic levels and changing the 
precession axis, disturb the scattering process and the polarization pattern thus 
betraying their presence, which is important for diagnostic purposes.

\begin{figure}[H]
\includegraphics*[scale=0.6, clip=true, trim=50 85 260 237]{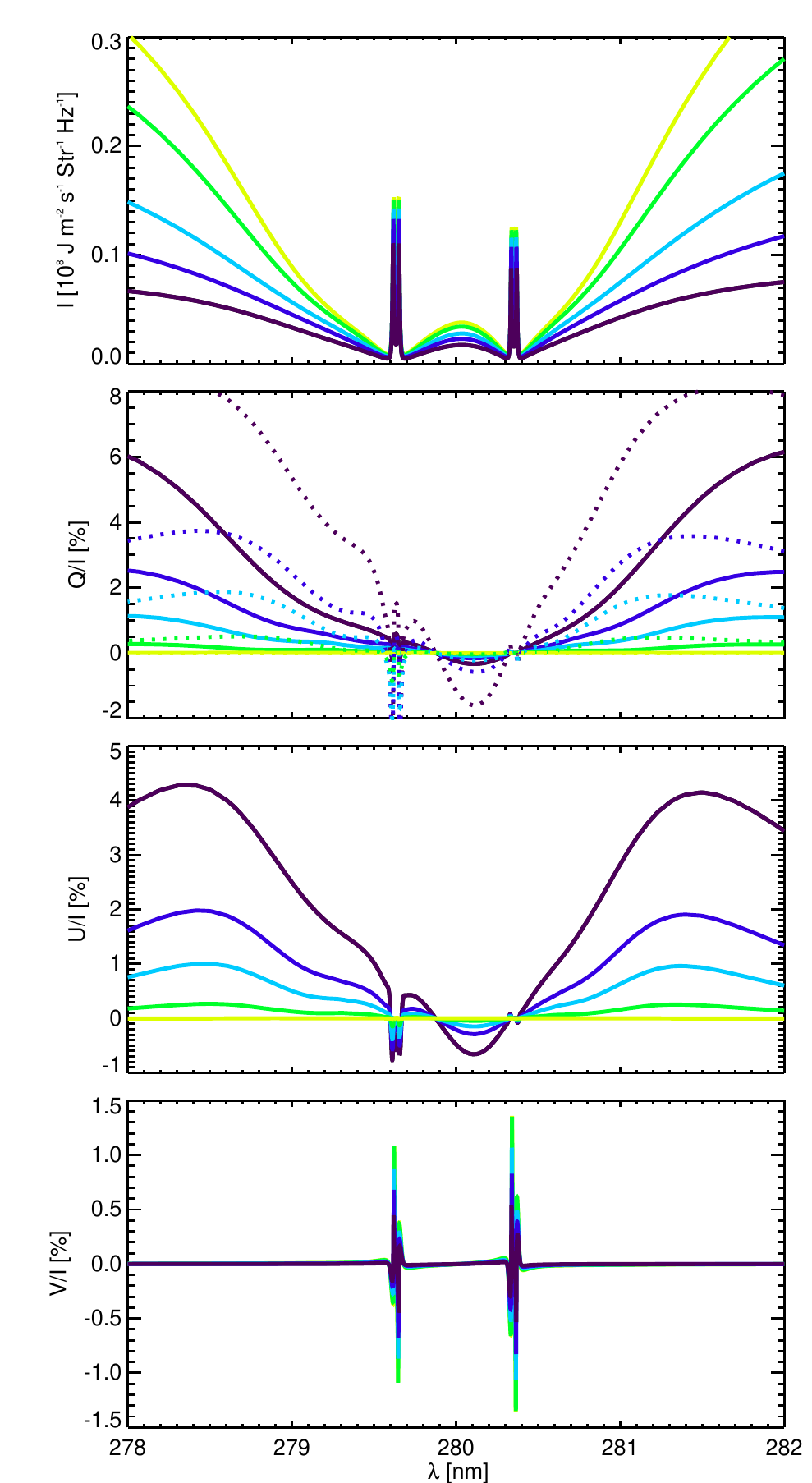}%
\includegraphics*[scale=0.6, clip=true, trim=50 85 200 237]{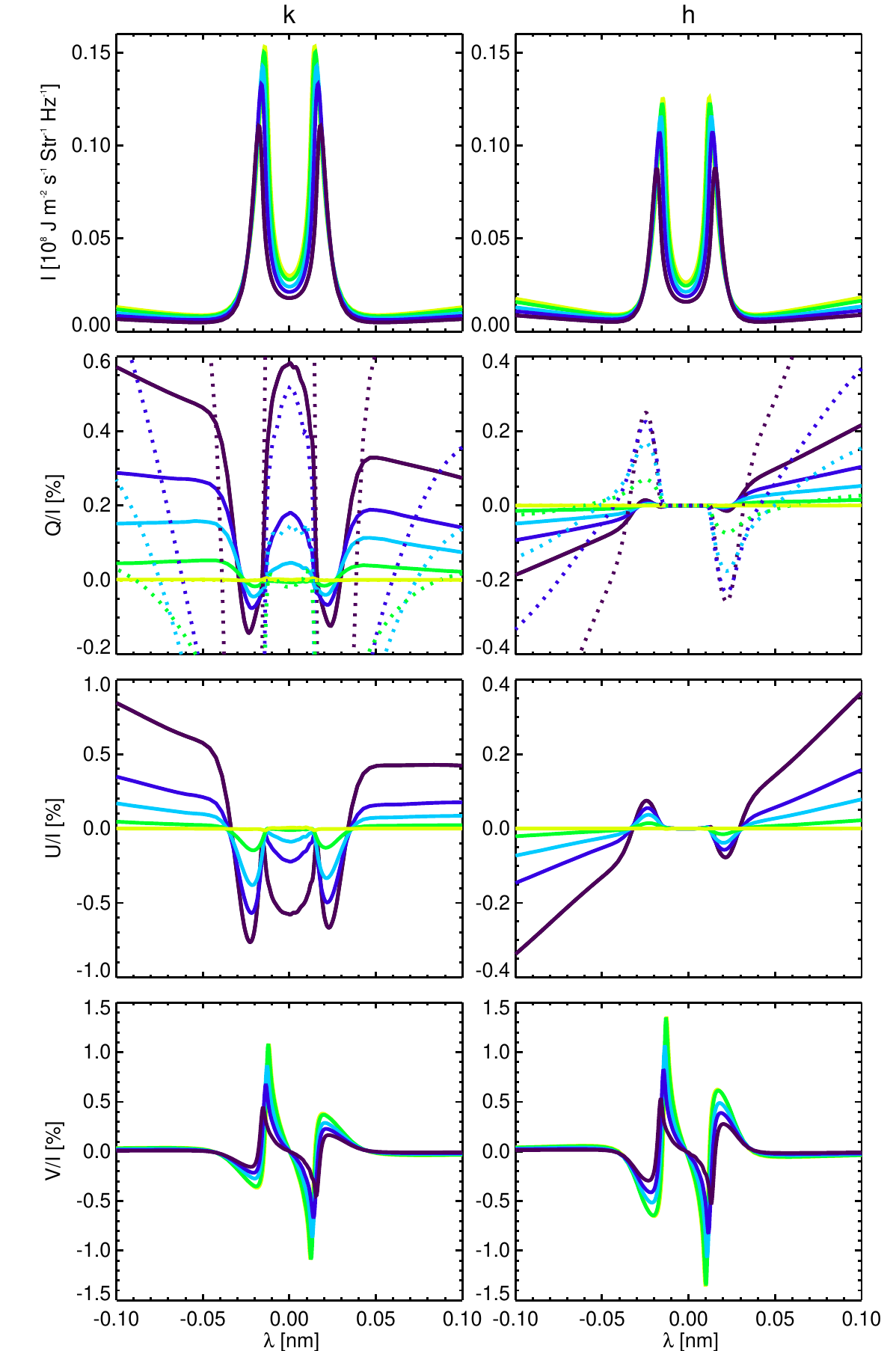}
\caption{\label{fig01}The linear polarization pattern (due to scattering) around the Mg~{\sc ii} k-h 
doublet calculated for a semiempirical solar model atmosphere \citep[FAL-C;][]{FAL93} 
is strongly modulated, even far from resonance, in the presence of
a magnetic field (here, $B=100$~G inclined $\theta_B=30^\circ$ with respect 
to the vertical and azimuth at $60^\circ$ to the line-of-sight---LOS). 
The circular polarization is well described by the magnetograph formula. 
Dotted lines show to the reference $B=0$ case.
Yellow, green, cyan, blue, violet, (or light to dark gray) 
for heliocentric angles $\theta$
with $\mu=\cos\theta=1$ (disc center), 0.8, 0.5, 0.3, 0.1, respectively.
}
\end{figure}

A general theory for describing light-matter interaction that accounts for PRD,
atomic polarization and coherence in a general multi-term atomic 
system, in the presence of arbitrary magnetic fields 
has been recently presented 
\citep{Casini+14, CasiniManso16, CasiniManso16b, Casini+17}.
We have developed a numerical polarization radiative transfer code 
that implements this theory to calculate 
the emergent Stokes profiles in planeparallel stellar atmospheres \citep{delPino+16}.
Its application to the formation of the Mg~{\sc ii} h and k doublet in the Sun 
has already provided a few notable surprises (Figure~\ref{fig01}).
Rotation (and depolarization) of linear scattering polarization is a
characteristic of the Hanle effect \citep{MoruzziStrumia91}. But it is also well-known  
that the Hanle effect mainly operates in the core spectral lines---here, the polarizable k-line 
(\citeauthor{Stenflo94} \citeyear{Stenflo94}; \citeauthor{LandiLandolfi04} \citeyear{LandiLandolfi04}, hereafter LL04).
Why, then, the remarkable rotation of the 
linear polarization pattern spanning several nm off-resonance in Figure~\ref{fig01}? 
Perhaps even more surprisingly, the reason is
magnetooptical (MO) effects 
\citep[][see also Alsina et al. 2016]{delPino+16}. 
We have checked this numerically 
but it is enlightening to integrate the radiative transfer equations 
for the Stokes parameters (LL04)
\begin{equation*}
\frac{d}{ds}
\left(\!\begin{array}{c} I \\ Q \\ U \\ V \end{array}\!\right) = 
- \left(\!\begin{array}{cccc} 
\eta_I & \eta_Q & \eta_U & \eta_V \\
\eta_Q & \eta_I & \rho_V & -\rho_U \\
\eta_U & -\rho_V & \eta_I & \rho_Q \\
\eta_V & \rho_U & -\rho_Q & \eta_I 
\end{array}\!\!\!\right)
\left(\!\begin{array}{c} I \\ Q \\ U \\ V \end{array}\!\right) 
+\left(\!\begin{array}{c} \varepsilon_I \\ \varepsilon_Q \\ \varepsilon_U \\ \varepsilon_V \end{array}\!\right) 
\end{equation*}
for a constant properties slab of total of optical thickness $\tau=\eta_I L$ ($L$ the geometrical thickness),
where $\eta_I$ is the absorption coefficient, $\eta_{Q, U, V}$ the dichroism coefficients,
$\rho_{Q, U, V}$ the MO coefficients, and $\epsilon_{I, Q, U, V}$ the emissivities.

\begin{figure}
\centering
\includegraphics*[scale=0.47, clip=true, trim=90 370 300 250]{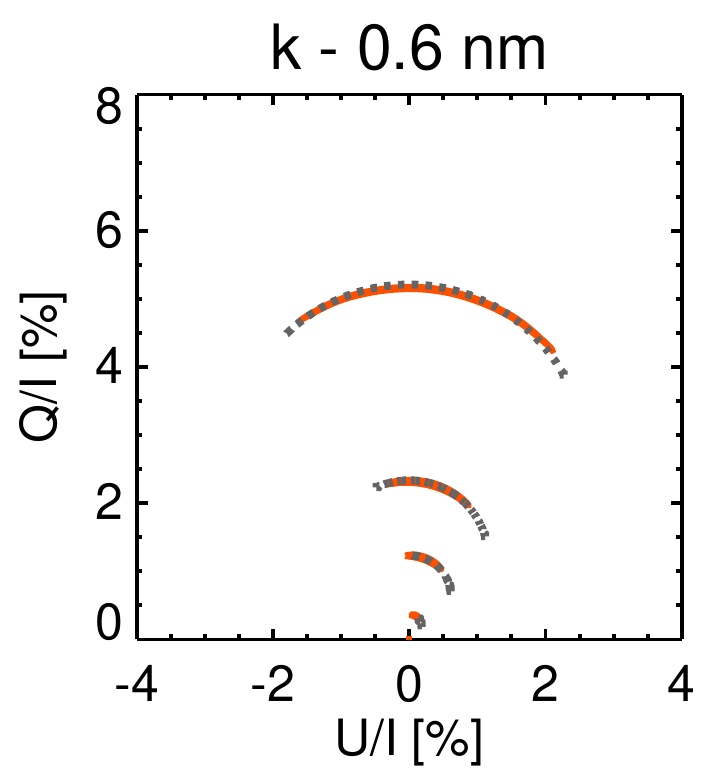}%
\includegraphics*[scale=0.47, clip=true, trim=100 370 295 250]{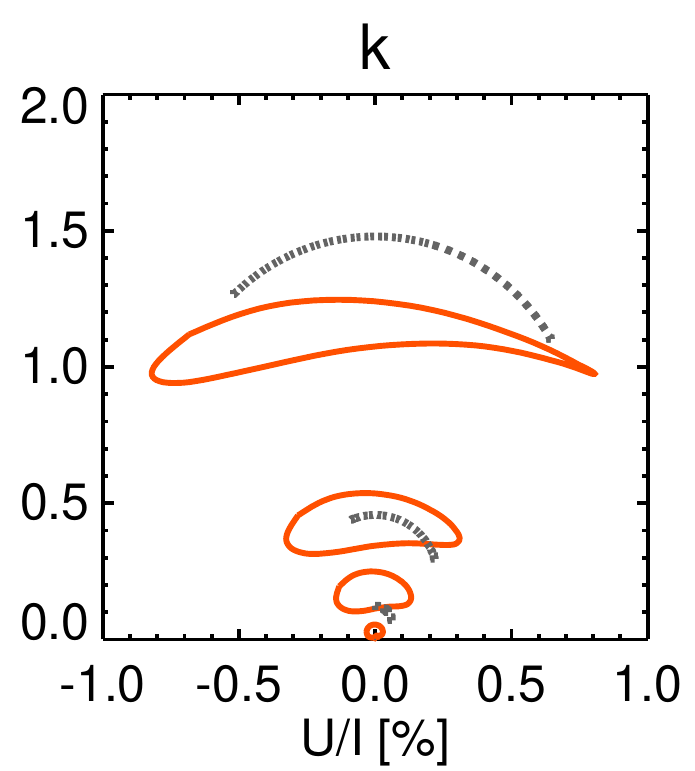}%
\includegraphics*[scale=0.47, clip=true, trim=100 370 295 250]{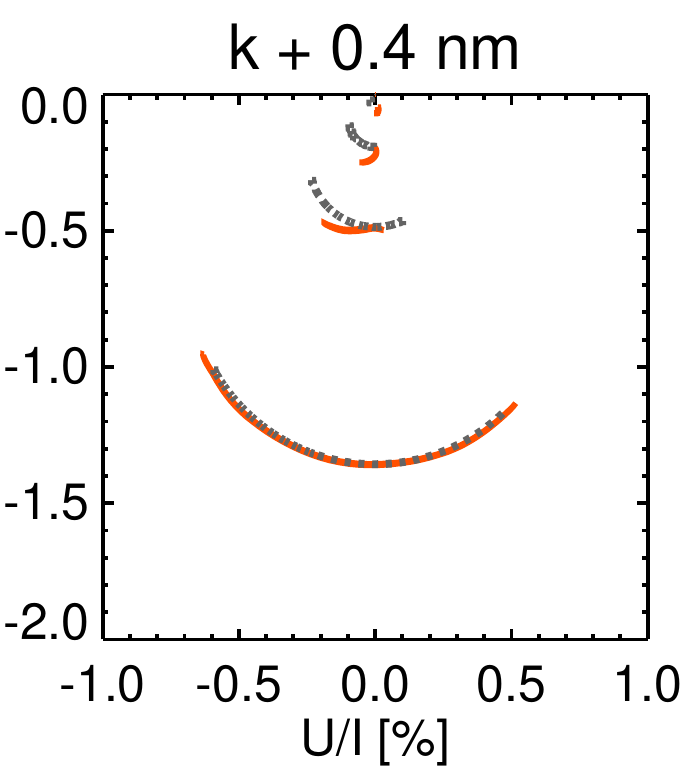}%
\includegraphics*[scale=0.47, clip=true, trim=110 370 300 250]{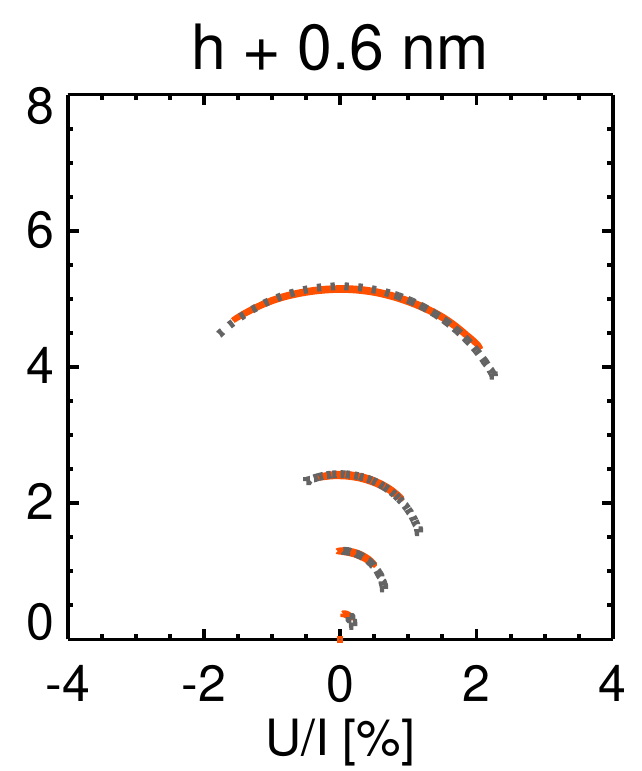}
\caption{\label{fig02}The Hanle effect in the core of the k-line (second panel), and the 
MO rotation of the scattering polarization pattern (in the far wings and in the negative branch 
between the h- and k-lines) look very different in the $Q$-$U$ plane. 
Solid lines: magnetic field $B=20$~G inclined $\theta_B=30^\circ$ 
to the vertical, arbitrary azimuth $0\leq \varphi_B < 2\pi$.
Four LOS are considered: $\mu=0.1$, the curve with the largest polarization in every panel,
and then, $0.3, 0.5, 0.8$, with decrasing polarization values. 
Dotted lines according to Equation~(\ref{eq04}), with 
$\rho=\alpha\cos\Theta_B=\alpha (\sqrt(1-\mu^2)\sin\theta_B\cos\varphi_B+\mu\cos\theta_B)$.
}
\end{figure}

For unpolarized lower levels, the $\rho$ and $\eta$ coefficients depend on the magnetic field
through the Zeeman splitting of the line profiles. Let $\Delta$ be the Zeeman splitting normalized to the line width.
In the weak field limit, $\rho_V$ and $\eta_V$ $\sim\Delta$, while
$\eta_{Q, U}$, $\rho_{Q, U}\sim\Delta^2$ (LL04).
We shall assume $\rho_V \gg \eta_Q, \eta_U, \rho_Q, \rho_U$ and proceed perturbatively.
We will further assume $\rho_V\gg\eta_V$ in the line wings.
Then, the emergent radiation assuming unpolarized illumination ($I_0$) is
\begin{subequations}\label{eq01}
\begin{align}
I=&\; I_0 {\rm e}^{-\tau} + \frac{\varepsilon_I}{\eta_I} (1-{\rm e}^{-\tau}) + \dots
\\
Q=&\; \frac{1-(c-\rho s) {\rm e}^{-\tau}}{1+\rho^2}
\frac{\epsilon_Q}{\eta_I}
- \frac{\rho-(\rho c + s) {\rm e}^{-\tau}}{1+\rho^2}
\frac{\epsilon_U}{\eta_I} + \dots
\\
U=&\; \frac{\rho-(\rho c + s) {\rm e}^{-\tau}}{1+\rho^2}
\frac{\epsilon_Q}{\eta_I}
+ \frac{1-(c-\rho s) {\rm e}^{-\tau}}{1+\rho^2}
\frac{\epsilon_U}{\eta_I} + \dots
\end{align}
\end{subequations}
where the dots stand for higher order terms including dichroism, 
$\rho=\rho_V/\eta_I$, $c=\cos(\tau\rho)$, and $s=\sin(\tau\rho)$.
In the optically thin limit ($\tau\ll 1$), 
\begin{equation}\label{eq02}
I =  I_0 +\tau \left(\frac{\varepsilon_I}{\eta_I}-I_0\right),
\qquad
Q= \tau \frac{\varepsilon_Q}{\eta_I} - \tau^2 \frac{\rho_V}{\eta_I} \frac{\varepsilon_U}{\eta_I},
\qquad
U= \tau \frac{\varepsilon_U}{\eta_I} + \tau^2 \frac{\rho_V}{\eta_I} \frac{\varepsilon_Q}{\eta_I};
\end{equation}
in the optically thick limit ($\tau\gg 1$),
\begin{equation}\label{eq03}
I= \frac{\varepsilon_I}{\eta_I},
\qquad
Q= \frac{1}{1+\rho^2}\frac{\varepsilon_Q}{\eta_I} - \frac{\rho}{1+\rho^2} \frac{\varepsilon_U}{\eta_I},
\qquad
U= \frac{\rho}{1+\rho^2}\frac{\varepsilon_Q}{\eta_I} + \frac{1}{1+\rho^2} \frac{\varepsilon_U}{\eta_I}.
\end{equation}
In LTE, the case in classical Zeeman diagnostics, 
the hierarchy above on dichroism and MO coefficients 
also implies that $\epsilon_V\gg\epsilon_Q, \epsilon_U$, and then, MO terms in
(\ref{eq01}) become of higher order to the (ellipsis) dichroism terms, which are governed by the 
(zeroth order) intensity. 
Thus, MO are second order in the optical depth (Equation~[\ref{eq02}]), and high order 
on the Zeeman splitting ($\rho_V\times\epsilon_{Q, U}\sim \Delta \times\Delta^2$).
For those reasons, the fundamental importance of MO effects was firstly understood 
from their characteristic signature in the core of spectral lines, 
in spectropolarimetric observations of sunspots 
\citep{Wittmann71, Landi79, LandolfiLandi82, SkumanichLites87}.

\begin{figure}
\includegraphics*[scale=0.6, clip=true, trim=10 370 60 100]{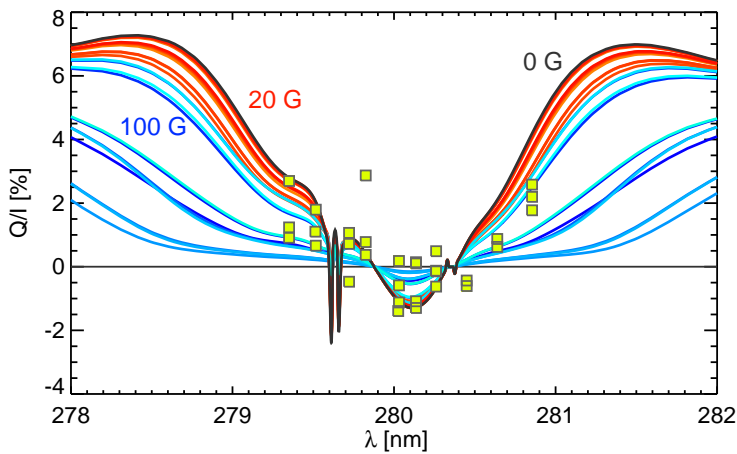}
\caption{\label{fig03}The recalibrated (substracting a $0.5\%$ offset) observations from 
UVSP/SMM reported by \citet[][yellow squares]{HenzeStenflo87} 
fall within the espected values from simulations. 
Here, synthetic $Q/I$ profiles from a FAL-C model at $\mu=0.15$ 
are shown for two magnetic field strengths:  
$B=20$~G (red) and 100~G (blue); the magnetic field inclination 
is $\theta_B=30^\circ$ and azimuths $0^\circ$, $30^\circ$, $60^\circ$, ...,
$330^\circ$, with respect to the LOS are shown 
($B=0$~G case in dark gray for reference).
The dispersion of the observations is not due to measurement errors (only), 
but to actual variations of the observed magnetic field.  
}
\end{figure}

Scattering polarization, however, is a zeroth order effect 
($\epsilon_{Q, U}\sim\Delta^0$)
and MO effects appear at lowest order on the perturbative analysis 
(Equations~[\ref{eq01}]).
Moreover, it is well known that inteference between $J$-levels of the same term
and PRD greatly enhance the scattering polarization patterns
around strong resonance lines 
\citep[][LL04]{Stenflo80, Auer+80, Stenflo96, BelluzziTrujillo11, Smitha+11, BelluzziTrujillo12}.
Surprisingly, only very recently has the role of MO effects described here been recognized in the 
magnetical modulation the linear polarization patterns around strong resonance lines 
\citep{delPino+16, Alsina+16}.

MO rotation in the wings is a transport phenomenon different from the Hanle effect
taking place in the core of the k-line (see Figure~2).
In the wings, we may neglect the variation of $\epsilon_{Q, U}$ with the magnetic field.
Then, from equations~(\ref{eq03}), 
\begin{equation}\label{eq04}
\frac{Q}{I}=\frac{1}{1+\rho^2} \left(\frac{Q}{I}\right)_0, \qquad 
\frac{U}{I}=\frac{\rho}{1+\rho^2} \left(\frac{Q}{I}\right)_0, 
\end{equation}
where the 0 subindex refers to the non magnetic case.
As $\rho$ varies (with field strength or geometry), equations~(\ref{eq04}) describe an arc
of a circle centered at $((Q/I)_0, (U/I)_0)$ in the $Q$-$U$ plane (cf. panels a, c, and d in Figure~\ref{fig02}).

\begin{figure}
\centering
\includegraphics*[scale=1.1, clip=true, trim=50 590 400 100]{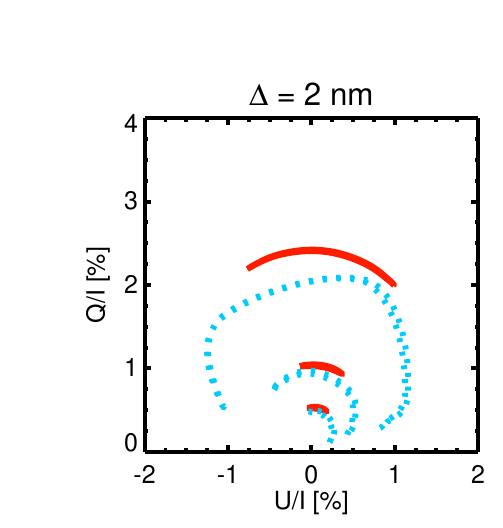}%
\includegraphics*[scale=1.1, clip=true, trim=50 590 400 100]{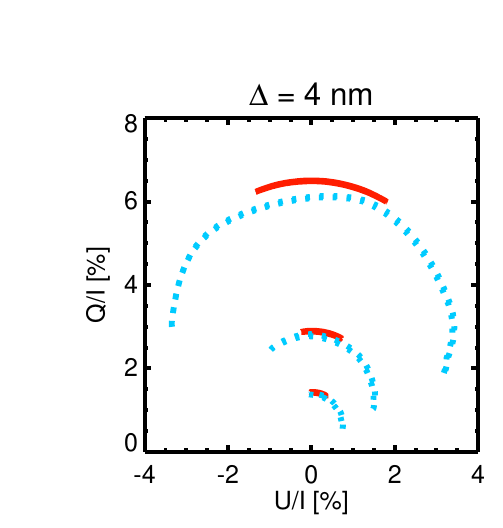}
\caption{\label{fig04}Broadband (passband $\Delta$) linear polarization 
is strongly modulated by the magnetic field geometry. 
Each curve corresponds to a fixed magnetic field inclination ($\theta_B=30^\circ$),
strength (solid lines: 20~G; dotted lines: 100G), LOS ($\mu=0.1$), 
and azimith $0\leq \varphi_B<2\pi$.
Larger polarization values are obtained with $\Delta=4$~nm which 
captures the polarization maxima on the wings at $\sim 1$~nm off-resonance (cf Figure 1). 
Cancellation due to the negative polarization branch between the $H$ and $K$ lines 
is more noticeable with smaller $\Delta$.
}
\end{figure}

Spectropolarimetric observations of 
linear polarization in the Mg~{\sc ii} h-k lines close to the solar limb 
were obtained by \citeauthor{HenzeStenflo87} (\citeyear{HenzeStenflo87}; see Figure~\ref{fig03})
using the Ultraviolet Spectrometer and Polarimeter \citep[UVSP;][]{Woodgate+80}
onboard the Solar Maximum Mission \citep[SMM; see][]{Strong+99}.
They confirmed the presence of strong polarization parallel to the limb in the wings 
of the lines (blue of the k-line, red of the h-line) due to scattering, 
but the existence of a negative (radial) polarization branch between the lines remained uncertain.
We have reanalysed the data applying an on-flight correction similar to  
the procedure of \citet{Giono+17}.
Observations at disk center with low spatio-temporal resolution are expected to be unpolarized 
due to symmetry reasons. 
However, the spatially and spectrally averaged 
UVSP/SMM observations at disk center show a residual $Q/I$ offset $\sim 0.5\%$.
When this spurious signal is removed from all the data (Figure~\ref{fig03}), the 
corrected data nicely fall between the 0~G profile and the $Q=0$ axis and in particular, 
the negative branch is confirmed. 
In fact, according to the MO theory explained here, the data points should not 
fall along the pure scattering profile ($B=0$~G line in Figure~\ref{fig03}) but rather fluctuate 
due to different magnetic field configurations likely present in the observed areas. 
The variability in the observations is therefore not (only) due to noise, but due to actual fluctuations of
the solar magnetic field. 
A reanalysis of that data set, including the observed (unpublished) $U/I$ pattern, would be of great interest.

The large polarization on the wide extended wings and their sensitivity to 
the magnetic field \citep{delPino+16} suggests the prospect of broadband polarimetry (BBP).
Interestingly, contrary to what is often the rule in polarimetry,
the larger the passband, the largest the signal (Figure~\ref{fig04}).
That is because the negative branch (which leads to cancellations) and the lower polarization  
close to the lines are more than compensated by the strong polarization $\sim 1$~nm away 
from the lines.
This comes at the price; the wings form deeper in the atmosphere than the 
high chromosphere probed by the core of the h and k lines.
In numerical experiments we find that if the atmosphere is unmagnetized below 1~Mm then the MO modulation 
occurs only in the close neigborhood of the lines, the pattern away and between them 
remains unafected. 
Yet, we can think of at least two interesting applications for BBP.
It offers the possibility of polarimetric imaging of the magnetic
field in the lower chromosphere and photosphere, for example, from context or slit-jaw
images which would constrain and facilitate the interpretation of spectropolarimetric
observations. 
Then, study of stellar magnetic fields (and rotation) would greatly benefit from BBP at this wavelengths.

\begin{figure}[H]
\centering
\includegraphics*[scale=0.6, clip=true, trim=0 0 10 10 ]{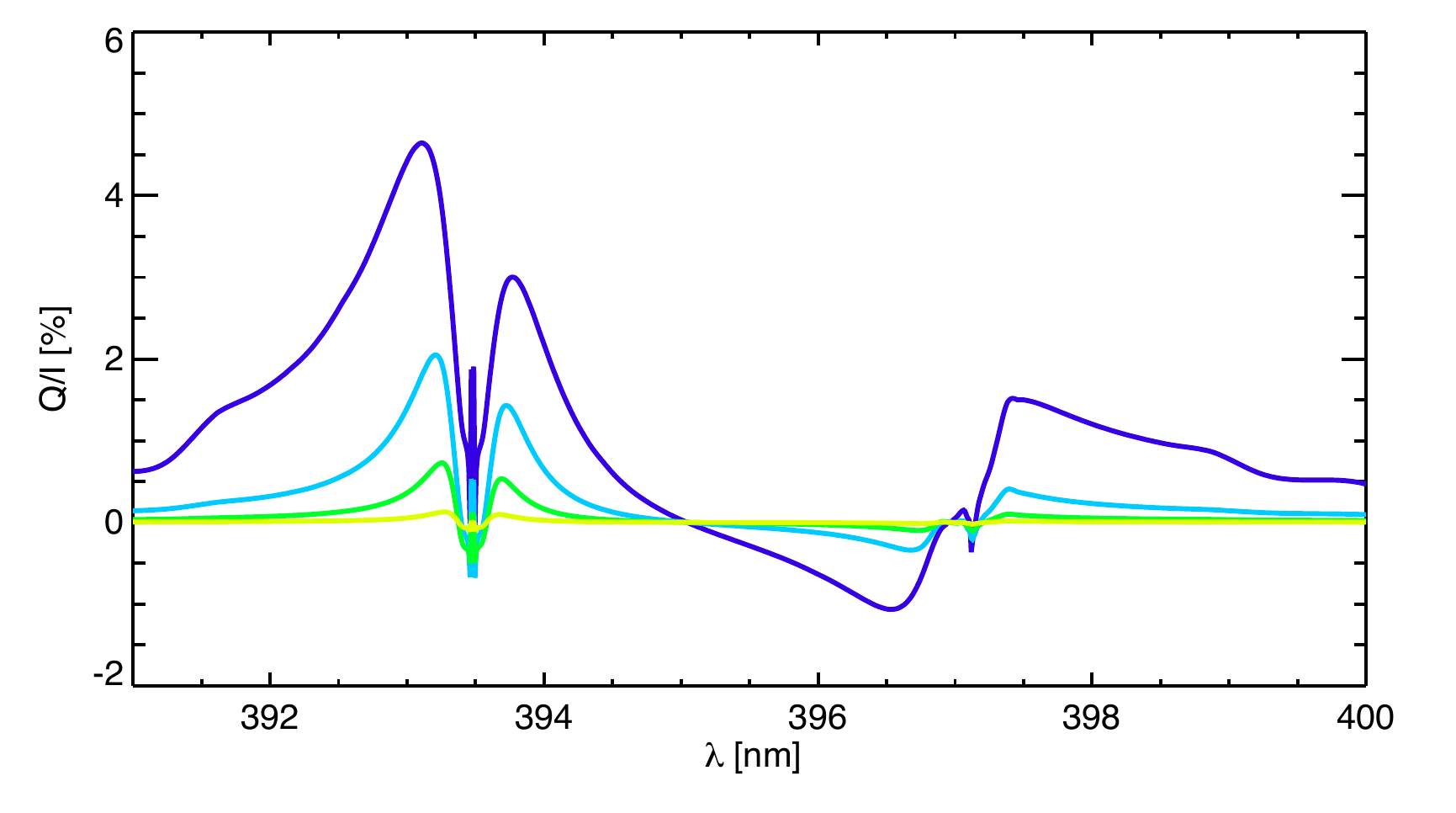}
\includegraphics*[scale=0.65, clip=true, trim=0 0 10 10 ]{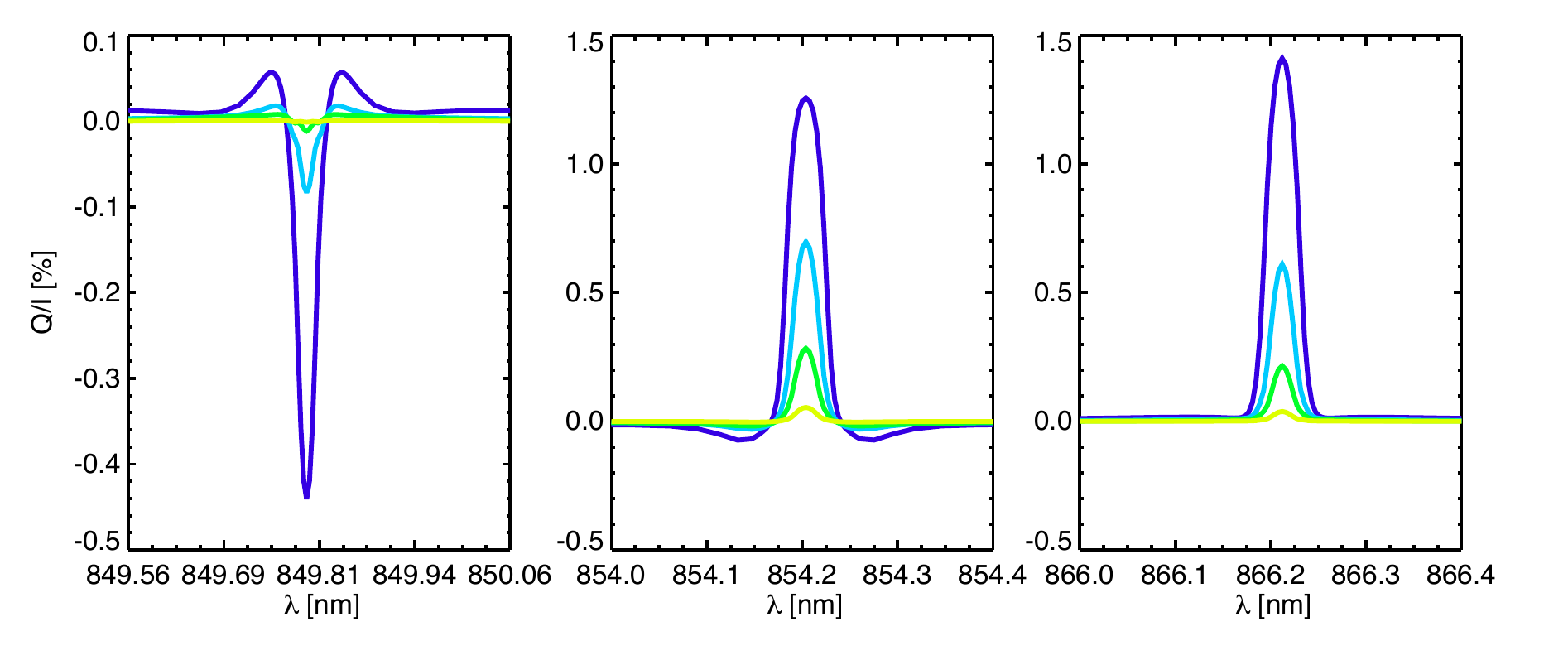}
\caption{The grand linear polarization pattern of the Ca~{\sc ii} $H$-$K$ 
doublet (upper panel) is largely due to scattering and PRD effects;
the scattering polarization signal in the 866.2 and 854.2~nm lines of the IR
triplet (lower panels) is fundamentally due to differential polarization absorption
(dichroism) by the aligned metastable ${}^2\!D_{3/2,5/2}$ levels.
The Ca~{\sc ii} $H$-$K$ lines and the IR triplet share the same upper term (${}^2\!P$).
Hence, PRD and atomic polarization have been simultaneously considered for a 
consistent description of scattering polarization in such $\Lambda$-type atom.
$Q/I$ calculated in a FAL-C solar model atmosphere at different
heliocentric angles ($\mu=0.1, 0.3, 0.5, 0.8$, for violet, cyan, green, yellow---or dark
to light gray---lines, respectively).
}
\end{figure}

In alkaline earth ions heavier than Mg, the first excited shell above the ground $s$ level
is not a $p$ but a $d$ electronic shell.
Consequently, the lowest (five) energy levels become a $\Lambda$-type system---a doublet
analogous to the Mg~{\sc ii} h and k lines, and a triplet between the doublets upper levels 
and the metastable $D$ term. 
Notably, in the case of Ca~{\sc ii}, both the H and K doublet and the IR triplet form in the
chromosphere, and both show remarkable scattering polarization patterns; 
the former dominated by PRD and level interference \citep{Stenflo80}, the latter
by dichroism from lower-level atomic polarization \citep{MansoTrujillo03}.
A consistent treatment of the relevant processes in this multilevel system 
is done for the first time in Figure~5. 
The role of the magnetic field coupling doublet and triplet systems cannot be discussed in this short note.


\end{document}